\documentclass[journal,twocolumn]{IEEEtran}
\usepackage{amssymb}
\usepackage{amsfonts}
\usepackage{amsmath}
\usepackage{algorithm}
\usepackage{algorithmic}
\usepackage{multicol}
\usepackage{amsmath}
\usepackage{subfigure}
\usepackage{graphicx}

\usepackage{graphicx,subfigure,amsmath,amssymb,cite}

\usepackage[dvips]{color}
\usepackage{float}
\ifCLASSINFOpdf
\else
\fi
\begin{document}

\title{Low-Complexity Linear Precoding for Secure Spatial Modulation }

\author{Feng Shu, \IEEEmembership{Member,~IEEE}, Zhengwang Wang, Shihao Yan, \IEEEmembership{Member,~IEEE}, \\Xiaobo Zhou, Jun Li, \IEEEmembership{Senior Member,~IEEE}, Xiangyun Zhou, \IEEEmembership{Senior Member,~IEEE}
\thanks{This work was in part supported by the National Natural Science Foundation of China (Grant Nos. 61771244, 61472190, 61702258, 61501238, and 61602245), the China Postdoctoral Science Foundation (2016M591852), the Postdoctoral research funding program of Jiangsu Province(1601257C), the Natural Science Foundation of Jiangsu Province (Grants No. BK20150791), and the open research fund of National Mobile Communications Research Laboratory, Southeast University, China (No.2013D02).}
\thanks{F. Shu, Z. Wang, X. Zhou, and J. Li are with School of Electronic and Optical Engineering, Nanjing University of Science and Technology, Nanjing, 210094, China. F. Shu is also with the College of Computer and Information Sciences, Fujian Agriculture and Forestry University, Fuzhou 350002, China.}
\thanks{S. Yan is with the School of Engineering, Macquarie University, Sydney, NSW 2109, Australia (e-mail: shihao.yan@mq.edu.au.)}
\thanks{X. Zhou is with Research School of Engineering, The Australian National University, ACT 2601, Australia (e-mail:  xiangyun.zhou@anu.edu.au.)}
}
\maketitle

% As a general rule, do not put math, special symbols or citations in the abstract or keywords.
\begin{abstract}
In this work, we investigate linear precoding for secure spatial modulation. With secure spatial modulation, the achievable secrecy rate does not have an easy-to-compute mathematical expression, and hence, has to be evaluated numerically, which leads to high complexity in the optimal precoder design. To address this issue, an accurate and analytical approximation of the secrecy rate is derived in this work. Using this approximation as the objective function, two low-complexity linear precoding methods based on gradient descend (GD) and successive convex approximation (SCA) are proposed. The GD-based method has much lower complexity but usually converges to a local optimum. On the other hand, the SCA-based method uses semi-definite relaxation to deal with the non-convexity in the precoder optimization problem and achieves near-optimal solution. Compared with the existing GD-based precoder design in the literature that directly uses the exact and numerically evaluated secrecy capacity as the objective function, the two proposed designs have significantly lower complexity. Our SCA-based design even achieves a higher secrecy rate than the existing GD-based design.
\end{abstract}

% Note that keywords are not normally used for peer-review papers.

\IEEEpeerreviewmaketitle

\section{Introduction}

\IEEEPARstart{S}{patial modulation (sm)}, proposed by Raed Y. Mesleh in \cite{Mesleh2008Spatial}, constitutes a promising multiple-input-multiple-output (MIMO) communication technology for future wireless communication systems. The basic idea of SM is to use both the indices of transmit antennas and constellation signals to carry a block of information bits \cite{Renzo2012Spatial,Jeganathan2008Spatial,di2014spatial}. Compared to single-input-single-output (SISO) systems, SM systems improve the overall spectral efficiency (SE) by the logarithm with base 2 of the number of transmit antennas~\cite{Mesleh2008Spatial}. Due to one antenna being active at any time instant in SM systems, the impacts of inter-channel interference (ICI) and inter-antenna synchronization (IAS) can be mitigated \cite{yang2015design}, and thus the practical implementation complexity at the transmitter and receiver is significantly reduced \cite{serafimovski2013practical}. Compared to space-time block coding (STBC) and bell laboratories layered space-time (BLAST) architectures, the SM technique strikes an attractive tradeoff between the SE and energy efficiency (EE), thus it is applicable to some future energy-efficient scenarios such as internet of things, 5G and beyond mobile networks. However, it is very likely that the confidential messages are intercepted by unintended receivers, due to the broadcast nature and openness of wireless channel. Therefore, it is very important and necessary to address security issues for such SM systems.

Recently, physical layer security on conventional MIMO systems has been widely investigated in \cite{Wyner1975,Shiu2011Physical,Wang2012Distributed,Zhao2016Physical,Shu2016Robust}, which exploits the uniqueness and time-varying characteristics of wireless channel to obtain secure transmissions against eavesdroppers. Particularly, the authors in \cite{hu2017artificial,Feng2018secure} exploited the direction modulation (DM) technique and random frequency diverse arrays to obtain secure and precise transmissions, where only desired users with predefined specific directions and distances can receive confidential messages. However, some works on physical layer security of conventional MIMO systems assumed that the input signals follow Gaussian distribution~\cite{Wu2017Secure}, which is not practical in SM systems due to the number of transmit antennas being finite. In fact, finite alphabet inputs should be considered in SM systems. In \cite{Guan2013Sec,Wang2015Secrecy,Liu2017Secure,Wu2015Secret,Wu2016Transmitter,Chen2016Secure,Aghdam2016Physical,shu2018two,xia2018as,Wang2016Spatial,Jiang2017Secrecy,yang2018mapping}, the authors investigated secure transmissions for SM systems from different aspects.

The authors in \cite{Guan2013Sec} derived the secrecy mutual information with finite alphabet input for a SM multiple-input-single-output (MISO) system, and proposed a precoding scheme to degrade the detection performance at an eavesdropper by decreasing the Euclidean distance at the eavesdropper. However, the precoding scheme is not applicable for SM-MIMO systems. Both \cite{Wang2015Secrecy} and \cite{Liu2017Secure} addressed the security of SM systems with the aid of artificial noise (AN), which was projected into the null space of legitimate channels. In \cite{Wang2015Secrecy}, the random AN signals were radiated at the transmitter and the secrecy rate (SR) was analyzed. In~\cite{Liu2017Secure}, a full-duplex receiver with self-interference cancelation capability was employed to transmit AN signals. Meanwhile, the precoding-aided spatial modulation (PSM) was generalized to provide secure transmissions for one legitimate user in \cite{Wu2015Secret,Wu2016Transmitter} or multiple legitimate users in \cite{Chen2016Secure}. The security of PSM was enhanced by constructing time-varying precoder \cite{Wu2015Secret,Chen2016Secure} or optimizing the precoder through jointly minimizing the receive power at eavesdropper and maximizing the receive power at desired user \cite{Wu2016Transmitter}. Unlike traditional SM systems, the PSM uses the index of receive antenna to carry information bits and activates all transmit antennas, which reduces the complexity at receiver but results in the issues of IAS and ICI at the transmitter.  In \cite{shu2018two,xia2018as}, the authors explored the schemes of transmit antenna selection to enhance the security of SM-MIMO systems. A new idea of redefining the mapping rules between the information bits and the indices of transmit antennas was proposed to achieve secure transmissions in \cite{Wang2016Spatial}, and then the authors in \cite{Jiang2017Secrecy,yang2018mapping} further improved the security by exploiting the knowledge of the legitimate channel state information (CSI) to rotate both the indices of the transmit antennas and the constellation symbols. In \cite{Wang2016Spatial,Jiang2017Secrecy,yang2018mapping}, the legitimate CSI actually can be regarded as an encryption key to encrypt the confidential messages. In other words, the scheme of rotating both the indices of the transmit antennas and the constellation symbols actually does not destroy the receive signals at eavesdroppers. Therefore, when the legitimate channel is slow fading channel, the key is invariant during a long coherence interval, and thus the eavesdroppers may decode the key by using the cryptanalysis or machine learning methods.

On the other hand, linear precoding for SM systems has received some researchers' attention in \cite{Lee2015Generalized,Strekalovsky2015Pre,Garcia2016Power,Yang2016Transmit,Cheng2018Low,Jin2015Linear}. The authors in \cite{Lee2015Generalized} proposed two design schemes of generalized precoder, maximum minimum distance (MMD) and guaranteed Euclidean distance (GED), and an iterative algorithm to acquire effective solutions. Then a low-complexity method with smaller numbers of iterations was proposed to solve the MMD and GED problems in \cite{Cheng2018Low}. Meanwhile, the authors in \cite{Strekalovsky2015Pre,Garcia2016Power} constructed a similar MMD problem for space shift keying (SSK) systems, where the optimization problem was efficiently solved using semi-definite relaxation (SDR) techniques. In \cite{Yang2016Transmit}, the authors introduced two design criterions, maximizing the minimum Euclidean distance and minimizing the bit error rate (BER), to optimize the diagonal linear precoding matrix. In \cite{Jin2015Linear}, a new precoding scheme was proposed to improve the mutual information for generalized spatial modulation (GSM) systems by converting the GSM systems into a virtual MIMO system and employing the extended ellipsoid algorithm. However, all the aforementioned works optimized the linear precoding vector to improve the BER performance regardless of security. In \cite{Aghdam2016Physical}, the authors investigated the design of linear precoding to maximize the actual SR (Max-SR) for secure SSK systems and the Max-SR based on gradient descend (GD) method (Max-SR-GD) was also proposed to solve the corresponding optimization problem, but the computational complexity is extremely high due to the requirement of a high computational amount to evaluate the actual SR.

Compared to \cite{Aghdam2016Physical}, we will focus on the design of linear precoding with lower complexity or higher performance in this work. Our main contributions are summarized as follows:
\begin{enumerate}
 \item Due to the absence of a closed-form expression for SR in secure SM systems with finite alphabet input, it is hard to design a practical feasible linear precoder of Max-SR  with lower-complexity. To reduce the computational complexity of Max-SR precoder, an approximated SR (ASR) is derived and presented, which is used as an optimization objective function. Numerical Monte-Carlo simulations show that the obtained ASR is very close to the actual SR for different numbers of transmit antennas and different types of modulation.
 \item By making use of the closed-form expression of ASR, the optimization problem of maximizing ASR (Max-ASR) is casted and proposed. To solve the optimization problem, a low-complexity precoder,  called Max-ASR-GD,  is proposed to iteratively solve the problem by GD algorithm. It is well-known that the GD method can usually converge to a locally optimal solution. Our simulation results also show that the proposed Max-ASR-GD achieves a slightly lower SR than the Max-SR-GD proposed in \cite{Aghdam2016Physical} but with a significantly lower complexity.
 \item To further improve the SR performance in the optimization problem of Max-ASR,  which is a typical non-convex quadratically constrained quadratic programming (QCQP) problem and  NP-hard in general, its objective function is first relaxed and represented as the difference between two convex functions  by using SDR techniques. Then, the successive convex approximation (SCA) method is employed to iteratively solve the convex subproblems and obtain an approximately optimal solution. Finally, the proposed Max-ASR-SCA is proved to be convergent. From simulation results, it follows that the proposed Max-ASR-SCA outperforms the Max-SR-GD method and Max-ASR-GD in terms of achieving a higher SR. More importantly, the number of iterations of the proposed Max-ASR-SCA is smaller than those of Max-SR-GD and Max-ASR-GD.
\end{enumerate}

The remainder of this paper is organized as follows. In Section II, the system model of the considered secure SM-MIMO is presented, and the optimization problem of Max-ASR is formulated in Section III. In Section IV, two precoding methods, Max-ASR-GD and Max-ASR-SCA, are proposed to solve the optimization problem, and their computational complexities are analyzed. Numerical results are presented in Section V. Finally, we make our conclusions in Section VI.

\textit{Notation:} Matrices, vectors, and scalars are denoted by letters of bold upper case, bold lower case, and lower case, respectively. Signs $(\cdot)^{-1}$ and $(\cdot)^H$ denote inverse and conjugate transpose, respectively. Notation $\mathbb{E}\{\cdot\}$ stands for the expectation operation. $\textbf{I}_N$ denotes the $N\times N$ identity matrix, and $\mathrm{tr}(\cdot)$ denotes matrix trace.

\section{System Model}

\subsection{Secure Spatial Modulation}
\begin{figure}
\centering
{\includegraphics[width=0.38\textwidth]{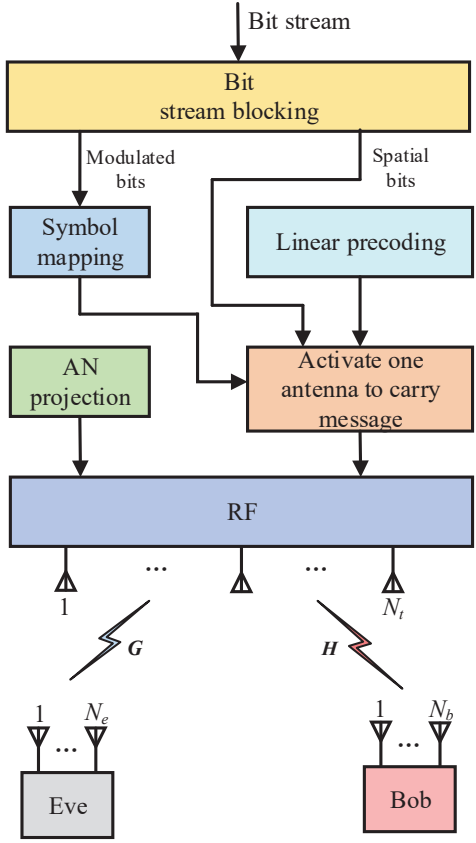}}\\
\caption{System model of linear precoding schemes for secure SM system}\label{system}
\end{figure}

A typical secure SM-MIMO system model is shown in Fig.~\ref{system}. In this system, there are a transmitter (Alice) with $N_t$ transmit antennas, a legitimate receiver (Bob) with $N_b$ receive antennas, and an eavesdropper (Eve) with $N_e$ receive antennas. Alice activates one of her $N_t$ transmit antennas to emit $M$-ary amplitude phase modulation (APM) symbol, and the index of activated antenna is also used to carry information. As a result, the SE is ${\log _2}{MN_t}$ bits per channel use (bpcu).

To enhance the security of such a SM system, a linear precoder and AN projection at transmitter is adopted to guarantee the secure transmission against Eve's eavesdropping. Similar to \cite{Wang2015Secrecy}, the transmitted baseband signal at Alice may be constructed as follows
\begin{align}\label{tran-sign}
\mathbf{x} =\sqrt{P_1}\mathbf{V}\mathbf{s}_{n,m} + \sqrt{P_2}\mathbf{T}_{AN}\mathbf{n},
\end{align}
where $\mathbf{V}\in \mathbb{C}^{N_t\times N_t}$ denotes the linear precoding matrix, and $\mathbf{T}_{AN}\in \mathbb{C}^{N_t\times N_t}$ is the AN projecting matrix. $P_1$ and $P_2$ are the power of confidential signal and AN, respectively, with power constraint $P_1+P_2 \leq P_t$ holding, where $P_t$ is the total transmit power. $\mathbf{n}\in \mathbb{C}^{N_t\times 1}$ is the random AN vector following standard complex Gaussian distribution $\mathcal{CN}(0,\mathbf{I}_{N_t})$. $\mathbf{s}_{n,m}=\mathbf{e}_ns_m$, $n\in[1,N_t]$, $m\in[1,M]$, is the input symbol vector. $s_m$ is the normalized input symbol, with $\mathbb{E}(|s_m|^2)=1$, which is drawn equiprobably from discrete $M$-ary constellation, and $\mathbf{e}_n$ is the $n$th column of identity matrix $\mathbf{I}_{N_t}$.

It is worth noting that the linear precoding for SM-MIMO systems is different from conventional MIMO systems. In SM-MIMO systems, only one antenna is activated to transmit modulated symbol, so the linear precoding matrix is a diagonal matrix, i.e., $\mathbf{V}=\rm{diag}(\mathbf{v})$, where $\mathbf{v}\in \mathbb{C}^{N_t\times 1}$ stands for the linear precoding vector. Moreover, the elements of linear precoding vector $\mathbf{v}$ cannot be active simultaneously. In other words, only the element corresponding to the activated transmit antenna is active during each symbol interval \cite{Lee2015Generalized}.

Let $\mathbf{H}\in \mathbb{C}^{N_b\times N_t}$ and $\mathbf{G}\in \mathbb{C}^{N_e\times N_t}$ denote the complex channel matrices corresponding to the legitimate channel and eavesdropping channel, respectively. This paper assumes that Alice has the perfect CSIs of both channels, which may be true for active Eve \cite{Aghdam2016Physical}. Meanwhile, Bob and Eve attain their own CSIs through pilot-assisted channel estimation, respectively. It is assumed that channel matrices $\mathbf{H}$ and $\mathbf{G}$ both are flat Rayleigh fading with each element being the standard complex-Gaussian distribution  $\mathcal{CN}(0,1)$ and keep constant during each coherence time. Accordingly, the receive signals at Bob and Eve are respectively stated as follows
\begin{align}\label{rece-bob}
{\mathbf{y}_b}&={\mathbf{H}}\mathbf{x} + {\mathbf{n}_b} \nonumber\\&= \sqrt {{P_1}}{\mathbf{H}}{\mathbf{V}}\mathbf{s}_{n,m} + \sqrt {{P_2}} {\mathbf{H}}{\mathbf{T}_{AN}}\mathbf{n} + {\mathbf{n}_b},
\end{align}
\begin{align}\label{rece-eve}
{\mathbf{y}_e}&={\mathbf{G}}\mathbf{x} + {\mathbf{n}_e} \nonumber\\&= \sqrt {{P_1}}{\mathbf{G}}{\mathbf{V}}\mathbf{s}_{n,m} + \sqrt {{P_2}} {\mathbf{G}}{\mathbf{T}_{AN}}\mathbf{n} + {\mathbf{n}_e},
\end{align}
where $\mathbf{n}_b$ and $\mathbf{n}_e$ are the additive white Gaussian noise (AWGN) vectors at Bob and Eve with obeying $\mathcal{CN}(0,\sigma_b^2\mathbf{I}_{N_b})$ and $\mathcal{CN}(0,\sigma_e^2\mathbf{I}_{N_e})$, respectively.

Consequently, with the knowledge of $\mathbf{H}$ and $\mathbf{V}$, Bob can employ the maximum likelihood detector (MLD) as follows
\begin{align}\label{ml-bob}
\left[ {\hat n,\hat m} \right] = \mathop {\arg \min }\limits_{n \in [1,{N_t}],m \in \left[ {1,M} \right]} {\left\| {{\mathbf{y}_b} - \sqrt {{P_1}}\mathbf{H}{\mathbf{V}}{\mathbf{s}_{n,m}}} \right\|^2}.
\end{align}

Here, we consider the worst case when Eve knows the linear precoding matrix $\mathbf{V}$, and thus Eve may also carry out MLD as follows
\begin{align}\label{ml-eve}
\left[ {\hat n,\hat m} \right] = \mathop {\arg \min }\limits_{n \in [1,{N_t}],m \in \left[ {1,M} \right]} {\left\| {{\mathbf{y}_e} - \sqrt {{P_1}}\mathbf{G}{\mathbf{V}}{\mathbf{s}_{n,m}}} \right\|^2}.
\end{align}
However, the terms of $\sqrt {{P_2}} {\mathbf{H}}{\mathbf{T}_{AN}}\mathbf{n}$ and $\sqrt {{P_2}} {\mathbf{G}}{\mathbf{T}_{AN}}\mathbf{n}$ in (\ref{rece-bob}) and (\ref{rece-eve}) is time-varying interference, which will seriously degrade their detecting performance. In what follows, we will derive an expression of SR with respect to $\mathbf{T}_{AN}$ and $\mathbf{V}$, then we project the AN signals into the null space of legitimate channel by designing the $\mathbf{P}_{AN}$. Finally, our main aim of this paper is to design a linear precoding matrix to improve SR performance.

\subsection{Secrecy Rate for Secure Spatial modulation}
Similar to \cite{Aghdam2016Physical}, we characterize the security by evaluating average SR defined as
\begin{align}\label{ave-sr}
{\bar R_s} = {\mathbb{E}_{\mathbf{H},\mathbf{G}}}\left( R_s \right),
\end{align}
where $R_s$ is the instantaneous SR. According to \cite{Wyner1975}, the SR for secure SM system is defined as follows
\begin{align}\label{Rs}
R_s={{{\left[ {I\left( {\mathbf{s};{\mathbf{y}_b}} \right) - I\left( {\mathbf{s};{\mathbf{y}_e}} \right)} \right]}^ + }},
\end{align}
where $[a]^+=\max\{a,0\}$. $I\left( {\mathbf{s};{\mathbf{y}_b}} \right)$ and $I\left( {\mathbf{s};{\mathbf{y}_e}} \right)$ denote the mutual information over legitimate and eavesdropping channels, respectively. However, due to the interference plus noise not being Gaussian distributed in (\ref{rece-bob}) and (\ref{rece-eve}), it is not straightforward to derive the expression of SR.

Similar to \cite{Wang2015Secrecy}, the interference plus noise at Bob is denoted as
\begin{align}\label{w_b}
\mathbf{w}_b=\sqrt {{P_2}}{\mathbf{H}}{\mathbf{T}_{AN}}\mathbf{n} + {\mathbf{n}_b},
\end{align}
then we utilize a linear whitening transformation matrix $\mathbf{Q}_b$ to convert $\mathbf{w}_b$ into a white noise vector
\begin{align}\label{wh-b}
\mathbf{n}_b'=\mathbf{Q}_b^{-1/2}\mathbf{w}_b,
\end{align}
where $\mathbf{Q}_b$ is the covariance matrix of $\mathbf{w}_b$ given by
\begin{align}\label{Q_b}
\mathbf{Q}_b={P_2}\mathbf{H}\mathbf{T}_{AN}\mathbf{T}_{AN}^H\mathbf{H}^H + \sigma_b^2\mathbf{I}_{N_b}.
\end{align}
Substituting (\ref{Q_b}) into (\ref{wh-b}) and considering that $\mathbf{n}$ and $\mathbf{n}_b$ are independent, we have $\mathbb{E}(\mathbf{n}_b')=0$ and $\mathbb{E}(\mathbf{n}_b'\mathbf{n}_b'^H)=\mathbf{I}_{N_b}$, thus $\mathbf{n}_b'\sim \mathcal{CN}(0,\mathbf{I}_{N_b})$ can be considered as an AWGN vector. By pre-multiplying the linear matrix $\mathbf{Q}_b$, the receiver signal in (\ref{rece-bob}) can be simplified as
\begin{align}\label{rece-bob-sim}
 \mathbf{y}_b'=\sqrt {{P_1}}\mathbf{Q}_b^{-1/2}{\mathbf{H}}{\mathbf{V}}\mathbf{s}_{n,m} + {\mathbf{n}_b'},
\end{align}
which yields, for a specifical $\mathbf{H}$, the conditional probability density function (PDF) of  $p\left( {{\mathbf{y}_b'}}|\mathbf{s}_{n,m} \right)$
\begin{align}\label{con-pdf}
p\left( {{\mathbf{y}_b'}\left| {{\mathbf{s}_{n,m}}} \right.} \right) = \frac{1}{{\pi}^{N_b}}\exp \left( {{ - {{\left\| {\mathbf{y}_b' -  \sqrt {{P_1}}\mathbf{Q}_b^{-1/2}\mathbf{H}{\mathbf{V}}{\mathbf{s}_{n,m}}} \right\|}^2}}}\right).
\end{align}
We assume that each antenna is activated equiprobably to transmit confidential message and each symbol $s_m$ is uniformly selected from $M$-ary constellation. Thus, the complex receive signal vector $\mathbf{y}_b$ at Bob obeys the distribution as follows
\begin{align}\label{pdf-bob}
p\left( {{\mathbf{y}_b'}} \right) = \frac{1}{{MN_t}}\sum\limits_{n = 1}^{N_t} {\sum\limits_{m=1}^M} p\left( {{\mathbf{y}_b'}\left| {{\mathbf{s}_{n,m}}} \right.} \right).
\end{align}
As per (\ref{con-pdf}) and (\ref{pdf-bob}), following the method given in \cite{Guan2013On}, the mutual information $I(\mathbf{s};\mathbf{y}_b')$ is written as
\begin{align}\label{mi-bob}
&I\left( {\mathbf{s};{\mathbf{y}_b'}} \right)= \log _2^{MN_t} - \frac{1}{{MN_t}}{\sum\limits_{n = 1}^{N_t}}{\sum\limits_{m=1}^M} {\mathbb{E}_{\mathbf{n}_b'}} \nonumber
\\&{\left[ {{{\log }_2}\left( {\sum\limits_{n' = 1}^{N_t} {\sum\limits_{m' = 1}^M {\exp \left( {{{{\left\| \mathbf{n}_b' \right\|}^2} - {{\left\| {\boldsymbol{\alpha} _{n,m}^{n',m'} + \mathbf{n}_b'} \right\|}^2}}} \right)} } } \right)} \right]},
\end{align}
where
\begin{align}\label{dis_bob}
\boldsymbol{\alpha} _{n,m}^{n',m'}=\sqrt {{P_1}}\mathbf{Q}_b^{-1/2}{\mathbf{H}}{\mathbf{V}}(\mathbf{s}_{n,m}-\mathbf{s}_{n',m'}).
\end{align}
It should be noted that the $I(\mathbf{s};\mathbf{y}_b')$ is equivalent to $I(\mathbf{s};\mathbf{y}_b)$, since the transformation is linear \cite{Wang2015Secrecy}. Similar to (\ref{mi-bob}), we can derive the mutual information $I(\mathbf{s};\mathbf{y}_e)$ as follows
\begin{align}\label{mi-eve}
&I\left( {\mathbf{s};{\mathbf{y}_e}} \right)= \log _2^{MN_t} - \frac{1}{{MN_t}}{\sum\limits_{n = 1}^{N_t}}{\sum\limits_{m=1}^M} {\mathbb{E}_{\mathbf{n}_e'}} \nonumber
\\&{\left[ {{{\log }_2}\left( {\sum\limits_{n' = 1}^{N_t} {\sum\limits_{m' = 1}^M {\exp \left( {{{{\left\| \mathbf{n}_e' \right\|}^2} - {{\left\| {\boldsymbol{\delta} _{n,m}^{n',m'} + \mathbf{n}_e'} \right\|}^2}}} \right)} } } \right)} \right]},
\end{align}
with
\begin{align}\label{dis_eve}
\boldsymbol{\delta} _{n,m}^{n',m'}=\sqrt {{P_1}}\mathbf{Q}_e^{-1/2}{\mathbf{G}}{\mathbf{V}}(\mathbf{s}_{n,m}-\mathbf{s}_{n',m'}),
\end{align}
\begin{align}\label{whiten}
\mathbf{n}_e'=\mathbf{Q}_e^{-1/2}(\sqrt {{P_2}} {\mathbf{G}}{\mathbf{T}_{AN}}\mathbf{n} + {\mathbf{n}_e}),
\end{align}
where $\mathbf{Q}_e={P_2}\mathbf{G}\mathbf{T}_{AN}\mathbf{T}_{AN}^H\mathbf{G}^H + \sigma_e^2\mathbf{I}_{N_e}$ is the linear whitening transformation matrix at Eve.

\subsection{Design the AN Projecting Matrix}
In this paper, we consider the scenarios with $N_t\ge N_b$, thus the AN signals can be projected into the null space of legitimate channel. Here, we present a closed-form expression of AN projection matrix, which is different from the singular value decomposition (SVD) method in \cite{Wang2015Secrecy}. Accordingly, the $\mathbf{T}_{AN}$ can be directly calculated as follows \cite{Hu2016Robust}
\begin{equation}\label{p-an}
\mathbf{T}_{AN}=\frac{1}{\mu }\left[ {{\mathbf{I}_{{N_t}}} - {\mathbf{H}^H}{{\left( {\mathbf{H}{\mathbf{H}^H}} \right)}^{ - 1}}\mathbf{H}} \right],
\end{equation}
where $\mu=\parallel {{\mathbf{I}_{{N_t}}} - {\mathbf{H}^H}{{\left( {\mathbf{H}{\mathbf{H}^H}} \right)}^{ - 1}}\mathbf{H}} \parallel_{\rm{F}}$ is the normalized factor for satisfying ${\rm{tr}}(\mathbf{T}_{AN}\mathbf{T}_{AN}^H)=1$ with the $\parallel \mathbf{A} \parallel_{\rm{F}}$ representing Frobenius norm of a matrix $\mathbf{A}$. It can be seen from (\ref{p-an}) that $\mathbf{T}_{AN}$ is the projector of the null space of legitimate channel, thus we have the following equality
\begin{align}\label{ns}
\mathbf{H}\mathbf{T}_{AN}=\mathbf{0}, \mathbf{Q}_b=\sigma_b^2\mathbf{I}_{N_b}.
\end{align}

\section{Optimization Problem on the Linear Precoding Matrix}
In this section, we first present a optimization problem of Max-SR to optimize the linear precoding matrix. Then, a simple expression of ASR is developed to be the convenience of dealing with the original optimization problem of Max-SR. Furthermore, due to the linear precoding matrix $\mathbf{V}$ being diagonal, we turn to optimize the linear precoding vector $\mathbf{v}=\rm{Diag}(\mathbf{V})$ for sake of convenience.

\subsection{Maximizing the Secrecy Rate}
After designing the AN projecting matrix, we turn to optimize the linear precoding matrix by solving the following problem of Max-SR
\begin{subequations}\label{opt-pre-sr}
\begin{align}
&\mathop {{\rm{maximize}}}\limits_{\mathbf{V}} {\rm{~~~~}}{R_s}(\mathbf{V})
\\&{\rm{subject~to}}:~\frac{P_1{\rm{tr}}({\mathbf{V}}{\mathbf{V}}^H)} {N_t} \le P_t-P_2.
\end{align}
\end{subequations}
It is worth noting that the above optimization problem of designing the linear precoder matrix is intractable in general, and we need many evaluations for calculating the actual SR due to the absence of the closed-form expression of SR. Actually, the above optimization problem can be solved by using GD method as illustrated in \cite{Aghdam2016Physical}, but it has a high computational complexity. As we will see later, to reduce the computational complexity, we present a closed-form approximated expression of SR, which can be used as an effective metric to optimize the linear precoding matrix.

\subsection{Approximated Expression for Secrecy Rate}
In this subsection, we present the tight lower bounds on mutual information over legitimate and eavesdropping channels, respectively, and provide an accurate approximation to SR. Similar to \cite{Guan2013On} and \cite{Zeng2012Linear}, by using Jensen's inequality and the integrals of exponential function, we have the tight lower bound of $I\left( {\mathbf{s};{\mathbf{y}_b}} \right)$
\begin{align}\label{mi-bob-lb}
&I\left( {\mathbf{s};{\mathbf{y}_b}} \right)_{LB}= \log _2^{MN_t} - \nonumber\\&\frac{1}{{MN_t}}{\sum\limits_{n = 1}^{N_t}}{\sum\limits_{m=1}^M}{\left[ {{{\log }_2}\left( {\sum\limits_{n' = 1}^{N_t} {\sum\limits_{m' = 1}^M {\exp \left( {\frac{{-{{\left\| {\boldsymbol{\alpha} _{n,m}^{n',m'} } \right\|}^2}}}{{2}}} \right)} } } \right)} \right]}.
\end{align}
Similarly, the mutual information of eavesdropping channel can be lower bounded by
\begin{align}\label{mi-eve-lb}
&I\left( {\mathbf{s};{\mathbf{y}_e}} \right)_{LB}= \log _2^{MN_t} - \nonumber\\&\frac{1}{{MN_t}}{\sum\limits_{n = 1}^{N_t}}{\sum\limits_{m=1}^M}{\left[ {{{\log }_2}\left( {\sum\limits_{n' = 1}^{N_t} {\sum\limits_{m' = 1}^M {\exp \left( {\frac{{-{{\left\| {\boldsymbol{\delta} _{n,m}^{n',m'} } \right\|}^2}}}{{2}}} \right)} } } \right)} \right]}.
\end{align}
Then an accurate approximation of SR can be given by
\begin{align}\label{app-sr}
R_s'={{{\left[ {I\left( {\mathbf{s};{\mathbf{y}_b}} \right)_{LB} - I\left( {\mathbf{s};{\mathbf{y}_e}} \right)_{LB}} \right]}^ + }}.
\end{align}

To further make the optimization variable clear, the term ${{\left\| {\boldsymbol{\alpha} _{n,m}^{n',m'} } \right\|}^2}$ in (\ref{mi-bob-lb}) can be represented as \cite{Cheng2018Low}
\begin{subequations}\label{dis-norm}
\begin{align}
{\left\| {{\boldsymbol{\alpha }}_{n,m}^{n',m'}} \right\|^2}&=P_1\cdot{\left\| { \mathbf{Q}_b^{-1/2}{\mathbf{H}}{{\mathbf{V}}}({{\bf{s}}_{n,m}} - {{\bf{s}}_{n',m'}})} \right\|^{\rm{2}}}
\\&=P_1\cdot{\rm{tr}}\left( {\mathbf{Q}_b^{-H}{\mathbf{H}}{{\mathbf{V}}}\boldsymbol{\Delta}_{n,m}^{n',m'}{{\mathbf{V}}}^H{{\mathbf{H}}^H}} \right)
\\&=P_1\cdot{\mathbf{v}}^H\left( {{\mathbf{H}}^H\mathbf{Q}_b^{-H}{\mathbf{H}} \odot \boldsymbol{\Delta}_{n,m}^{n',m'}} \right){\mathbf{v}}
\\&=P_1\cdot{\mathbf{v}}^H\mathbf{B}_{n,m}^{n',m'}{\mathbf{v}},
\end{align}
\end{subequations}
where
\begin{align}\label{cor-bob}
&\boldsymbol{\Delta}_{n,m}^{n',m'} = ({{\bf{s}}_{n,m}} - {{\bf{s}}_{n',m'}}){({{\bf{s}}_{n,m}} - {{\bf{s}}_{n',m'}})^H},\\&
\mathbf{B}_{n,m}^{n',m'}=\left[ {({\mathbf{H}}^H\mathbf{Q}_b^{-H}{\mathbf{H}})\odot \boldsymbol{\Delta}_{n,m}^{n',m'}} \right].\label{B-mat}
\end{align}
Similarly, the term ${{\left\| {\boldsymbol{\delta} _{n,m}^{n',m'} } \right\|}^2}$ in (\ref{mi-eve-lb}) can be represented as
\begin{subequations}\label{cor-eve}
\begin{align}
{{\left\| {\boldsymbol{\delta} _{n,m}^{n',m'} } \right\|}^2}&= P_1\cdot{\mathbf{v}}^H\left[ \left({{\mathbf{G}}^H\mathbf{Q}_e^{-H}{\mathbf{G}}}\right) \odot \boldsymbol{\Delta}_{n,m}^{n',m'} \right]{\mathbf{v}}
\\&=P_1\cdot{\mathbf{v}}^H\mathbf{E}_{n,m}^{n',m'}{\mathbf{v}},
\end{align}
\end{subequations}
with
\begin{align}\label{E-mat}
\mathbf{E}_{n,m}^{n',m'}=\left[ \left({{\mathbf{G}}^H\mathbf{Q}_e^{-H}{\mathbf{G}}}\right) \odot \boldsymbol{\Delta}_{n,m}^{n',m'} \right],
\end{align}
where $\odot$ denotes the Hadamard product of two matrices. We note that (\ref{dis-norm}) and (\ref{cor-eve}) are achieved by utilizing the trace property, i.e., ${\rm{tr}}(\mathbf{A}\mathbf{B})={\rm{tr}}(\mathbf{B}\mathbf{A})$ for matrices $\mathbf{A}$ and $\mathbf{B}$.

\subsection{Maximizing the Approximated SR}
Making use of the above accurate approximation of SR, the Max-SR optimization problem in (\ref{opt-pre-sr}) can be converted into the following simplified problem
\begin{subequations}\label{opt-approx}
\begin{align}
&\mathop {{\rm{maximize}}}\limits_{\mathbf{v}} {\rm{~~~~}}{{R_s'}}\left( {{\mathbf{v}}} \right)
\\&{\rm{subject~to}}:~~{\rm{tr(}}{\mathbf{v}}{\mathbf{v}}^H{\rm{)}} \le N_t\label{opt-approx_b},
\end{align}
\end{subequations}
where ${{R_s'}}\left( {{\mathbf{v}}} \right)$ is given in (\ref{Rs-app}) at the top of next page, and the (\ref{opt-approx_b}) is obtained with $P_1+P_2\leq P_t$. The above Max-ASR can significantly reduce the complexity compared to Max-SR in (\ref{opt-pre-sr}) as shown in what follows. However, it can be seen from (\ref{Rs-app}) that the above optimization problem is a non-convex QCQP problem, thus it is NP-hard in general. In what follows, we present a lower-complexity GD method and propose the SCA method to address the above optimization problem.
\begin{figure*}
\begin{align}\label{Rs-app}
{{R_s'}}\left( {{\mathbf{v}}} \right) = \frac{1}{{MN_t}}{\sum\limits_{n = 1}^{N_t}}{\sum\limits_{m=1}^M} {\left[ {{{\log }_2}\left( {\sum\limits_{n' = 1}^{N_t} {\sum\limits_{m' = 1}^M {\exp \left( {\frac{{- P_1{\rm{tr}}( {\mathbf{v}^H}\mathbf{E}_{n,m}^{n',m'}{\mathbf{v}})}}{{2}}} \right)} } } \right)}-{{{\log }_2}\left( {\sum\limits_{n' = 1}^{N_t} {\sum\limits_{m' = 1}^M {\exp \left( {\frac{{- P_1{\rm{tr}}( {\mathbf{v}^H}\mathbf{B}_{n,m}^{n',m'}{\mathbf{v}})}}{{2}}} \right)} } } \right)} \right]}
\end{align}
\hrulefill
\end{figure*}

\section{Propose Low-Complexity Linear Precoding Schemes of Maximizing ASR}
To provide a rapid solution to the optimization problem in (\ref{opt-approx}), two low-complexity methods, Max-ASR-GD and Max-ASR-SCA, are presented in this section. For the former, similar to \cite{Aghdam2016Physical},  using a gradient vector of the closed-form ASR, the iterative GD algorithm can be employed to yield an extremely low-complexity solution to the optimization problem as given in (\ref{opt-approx}), which converges to a locally optimal solution. For the latter, the original problem is first relaxed to a difference of convex (DC) programming by using the SDR method, and the first-order Taylor approximation  expansion and SCA way  are combined to achieve an approximately optimal solution.

\subsection{Proposed Max-ASR-GD Method}
Due to the nonconvexity of the problem (\ref{opt-approx}), it is intractable to obtain a globally optimal solution in general. However, it is possible to employ the GD method, which iteratively searches for a locally optimal solution. The  Max-ASR-GD method firstly needs to calculate the gradient of $R_s'(\mathbf{v})$ with respect to $\mathbf{v}$. This gradient is directly shown in (\ref{grad}) at the top of the next page.
\begin{figure*}
\begin{align}\label{grad}
{\nabla _{{\mathbf{v}}}}{R'_s}\left( {{\mathbf{v}}} \right) = \frac{{{P_1}}}{{MN_t}}\sum\limits_{n = 1}^{N_t}{\sum\limits_{m = 1}^M} {\left[ {\frac{{\sum\limits_{n' = 1}^{N_t} {\sum\limits_{m' = 1}^M {\left[ {\exp \left( {\frac{{ - P_1\cdot{\mathbf{v}}^H\mathbf{B}_{n,m}^{n',m'}{\mathbf{v}}}}{2}} \right)\mathbf{B}_{n,m}^{n',m'}{\mathbf{v}}} \right]} } }}{{2{\kappa_b}\cdot{{\ln}2}}} - \frac{{\sum\limits_{n' = 1}^{N_t} {\sum\limits_{m' = 1}^M {\left[ {\exp \left( {\frac{{ - P_1\cdot{\mathbf{v}}^H\mathbf{E}_{n,m}^{n',m'}{\mathbf{v}}}}{2}} \right)\mathbf{E}_{n,m}^{n',m'}{\mathbf{v}}} \right]} } }}{{2{\kappa _e}\cdot{{\ln}2}}}} \right]}
\end{align}
\hrulefill
\end{figure*}
In (\ref{grad}),  $\kappa_b$ and $\kappa_e$ are defined as follows
\begin{align}\label{kb}
{\kappa _b} = \sum\limits_{n' = 1}^{N_t} {\sum\limits_{m' = 1}^M {\exp \left( {\frac{{ - P_1\cdot{\mathbf{v}}^H\mathbf{B}_{n,m}^{n',m'}{\mathbf{v}}}}{{2}}} \right)} },
\end{align}
\begin{align}\label{ke}
{\kappa _e} = \sum\limits_{n' = 1}^{N_t} {\sum\limits_{m' = 1}^M {\exp \left( {\frac{{ - P_1\cdot{\mathbf{v}}^H\mathbf{E}_{n,m}^{n',m'}{\mathbf{v}}}}{2}} \right)} }.
\end{align}

Based on the gradient of $R_s'(\mathbf{v})$ over $\mathbf{v}$ in (\ref{grad}) ,  we update the linear precoding vector in the following way
\begin{align}\label{update}
\mathbf{v}_{k+1}=\mathbf{v}_k+\mu{\nabla_{\mathbf{v}}}{R_{s}'(\mathbf{v}_k)},
\end{align}
where $\mu$ and $k$ are the step size and iterative index, respectively. It should be noted that the updated linear precoding vector should satisfy the power constraint, the power normalization procedure is written as
\begin{align}\label{norm}
\mathbf{v}=\sqrt{N_t/{\rm{tr}}(\mathbf{v}\mathbf{v}^H)}~\mathbf{v}.
\end{align}

The detailed procedure of the proposed Max-ASR-GD method is illustrated in Algorithm \ref{Alg1}. The algorithm iteratively searches for an optimal $\mathbf{p}$. It is guaranteed that this algorithm converges to a locally optimal solution.
\begin{algorithm}[http]
\caption{The proposed Max-ASR-GD method for solving problem (\ref{opt-approx})}\label{Alg1}
\begin{algorithmic}[1]
\STATE Initialize $\mathbf{v}_0$ with satisfying ${\rm{tr}}(\mathbf{v}_0\mathbf{v}_0^H)\leq N_t$, set the step size $\mu$, the minimum tolerance $\mu_{\rm{min}}$ and $k=0$
\STATE Calculate ${R_{s}'(\mathbf{v}_k})$ by using (\ref{Rs-app})
\STATE Calculate ${\nabla _{\mathbf{v}}{R_{s}'(\mathbf{v}_k)}}$ by using (\ref{grad})
\STATE If $\mu\geq \mu_{\rm{min}}$, goto step 5, otherwise stop and return $\mathbf{v}_k$
\STATE Calculate $\mathbf{v}_{k+1}$ by using (\ref{update}) and normalize $\mathbf{v}_{k+1}$ by using (\ref{norm})
\STATE Calculate ${R_{s}'(\mathbf{v}_{k+1})}$ by using (\ref{Rs-app})
\STATE If ${R_{s}'(\mathbf{v}_{k+1})}\geq{R_{s}'(\mathbf{v}_k)}$, then goto step 8, otherwise, let $\mu=\mu/2$ and goto step 4
\STATE Do $k=k+1$ goto step 3
\STATE Output the $\mathbf{v}_k$, then calculate the SR $R_s(\mathbf{v}_k)$ by using (\ref{mi-bob}) and (\ref{mi-eve})
\end{algorithmic}
\end{algorithm}

It should be noted that the authors in \cite{Aghdam2016Physical} proposed a Max-SR-GD method to solve an optimization problem similar to (\ref{opt-pre-sr}). However, the method Max-SR-GD in \cite{Aghdam2016Physical} is to optimize the linear precoding vector in secure SSK systems, and it can be steadily extended to secure SM systems. There exists two serious problems of facing Max-SR-GD: no closed-form expression for SR and high-complexity. No closed-form expression means that it requires complex computation to evaluate the actual value of SR and its gradient per iteration.

\subsection{Proposed Max-ASR-SCA Method}
The optimization problem (\ref{opt-approx}) is a non-convex QCQP problem, which motivates us to explore the SDR method. The SDR is a powerful and computationally efficient approximation technique, which is particularly applicable to non-convex QCQP problems. Many practical experiences have indicated that SDR is capable of providing near optimal approximations \cite{Luo2010Semidefinite}.

Let us introduce a new matrix variable $\mathbf{W}=\mathbf{v}\mathbf{v}^H$, then we obtain the equivalence of (\ref{opt-approx}) as follows
\begin{subequations}\label{sdp-opt}
\begin{align}
&\mathop {{\rm{maximize}}}\limits_{{\mathbf{W}}} ~~~~{{R_s'}}\left( {{\mathbf{W}}} \right)
\\&{\rm{subject~to}}:~~{\rm{tr}}(\mathbf{W}) \le N_t
\\&~~~~~~~~~~~~~~~~~{\mathbf{W}} \succeq 0
\\&~~~~~~~~~~~~~~~~~{\rm{rank}}(\mathbf{W})=1,
\end{align}
\end{subequations}
where $\mathbf{W}\succeq 0$ denotes that $\mathbf{W}$ is a semi-definite matrix. However, the above optimization problem still is intractable due to the existence of the non-convex rank-one constraint.

Actually, after removing the non-convex rank-one constraint in the above optimization problem, (\ref{sdp-opt}) is relaxed as
\begin{subequations}\label{sdr-opt}
\begin{align}
&\mathop {{\rm{maximize}}}\limits_{\mathbf{W}} ~~~~{{R_s'}}\left({\mathbf{W}}\right)
\\&{\rm{subject~to}}:~~{\rm{tr}}(\mathbf{W}) \le N_t
\\&~~~~~~~~~~~~~~~~~{\mathbf{W}} \succeq 0.
\end{align}
\end{subequations}
By dropping the rank-one constraint, the above SDR achieves a upper bound on the optimal solution to the primal problem (\ref{sdr-opt}). If the optimal solution $\mathbf{W}^*$ to (\ref{sdr-opt}) is rank-one, the optimal $\mathbf{V}^*$ is the eigenvector corresponding to the largest eigenvalue of $\mathbf{W}^*$. Otherwise, the randomization technique may be used to obtain approximative optimum. In what follows, we focus on solving the optimization problem in (\ref{sdr-opt}).

In (\ref{sdr-opt}), the two constraints are convex sets. But its objective function is in form of subtraction of two convex functions \cite{Convex}. In general, this difference is non-concave. Obviously, the objective function in (\ref{sdr-opt}) is also non-concave. It is well-known that  SCA is an efficient way to solve this non-convex DC programming problem. The algorithm begins with an initial feasible point, the non-convex objective function is approximated by a strictly convex around this point. The resulting convex problem is solved to obtain the feasible point for next iteration. The iteration is repeated until the stopping criterion is satisfied \cite{Mehanna2014Feasible}.

For convenience of presentation, we rewrite the objective function as
\begin{align}\label{re-obj}
{{R_s'}}\left( {\mathbf{W}} \right)=\frac{1}{{MN_t}}{\sum\limits_{n = 1}^{N_t}}{\sum\limits_{m=1}^M}\left[ {{{f_1}\left( {{\mathbf{W}}} \right) - {f_2}\left( {{\mathbf{W}}} \right)} }\right],
\end{align}
where ${f_1}\left( {{\mathbf{W}}} \right)$ and ${f_2}\left( {{\mathbf{W}}} \right)$ are defined as
\begin{align}\label{f-1}
{f_1}\left( {{\mathbf{W}}} \right)={{{\log }_2} {\sum\limits_{n'= 1}^{N_t} {\sum\limits_{m'= 1}^M {\exp \left( {\frac{{- {\rm{tr}}( {\mathbf{W}}\mathbf{E}_{n,m}^{n',m'})}}{{2}}} \right)} } }},
\end{align}
\begin{align}\label{f-2}
{f_2}\left( {{\mathbf{W}}} \right)={{{\log }_2} {\sum\limits_{n'= 1}^{N_t} {\sum\limits_{m'= 1}^M {\exp \left( {\frac{{- {\rm{tr}}( {\mathbf{W}}\mathbf{B}_{n,m}^{n',m'})}}{{2}}} \right)} } } }.
\end{align}
In (\ref{re-obj}), ${f_1}\left( {{\mathbf{W}}} \right)$ and ${f_2}\left( {{\mathbf{W}}} \right)$ are log-sum-exp functions, which are proved to be convex \cite{Convex}, as a result, ${R_s'}(\mathbf{W})$ is non-concave.

It is well-known that a linear function is both convex and concave. In the following, we employ the first-order Taylor series expansion around a feasible point $\mathbf{W}_0$ on ${f_1}\left( {{\mathbf{W}}}\right)$ to transform ${f_1}\left( {\mathbf{W}} \right)$ into a linear function. Due to $f_1(\mathbf{W})$ being convex and differentiable, the first-order Taylor approximation actually is a global underestimator of the function \cite{Convex}, i.e., the following inequality holds
\begin{align}\label{taylor}
{f_1}\left( {{\mathbf{W}}} \right) \ge {f_1}\left( {{\mathbf{W}_0}} \right) + {\rm{tr}}\left[ {\nabla{f_1}\left( {{\mathbf{W}_0}} \right)\left( {\mathbf{W} - {\mathbf{W}_0}} \right)} \right],
\end{align}
where $\nabla{f_1}\left( {{\mathbf{W}_0}} \right)$ is defined as the gradient of the function ${f_1}\left( {\mathbf{W}} \right)$ at the point $\mathbf{W}_0$, which is derived as
\begin{align}\label{grad-f1}
\nabla{f_1}\left( {{\mathbf{W}_0}} \right) = -
\frac{{\sum\limits_{n' = 1}^{N_t} {\sum\limits_{m' = 1}^M {\left[ {\exp \left( {\frac{{ - {\rm{tr}}\left( {{\mathbf{W}_0}\mathbf{E}_{n,m}^{n',m'}} \right)}}{2}} \right){\mathbf{E}_{n,m}^{n',m'}}^H} \right]} } }}{{2{{\ln2}}\sum\limits_{n' = 1}^{N_t} {\sum\limits_{m' = 1}^M {\exp \left( {\frac{{ - {\rm{tr}}\left( {{\mathbf{W}_0}{\mathbf{E}_{n,m}^{n',m'}}} \right)}}{2}} \right)} } }}.
\end{align}
Then, given a fixed $\mathbf{W}_{k-1}$, the problem (\ref{sdr-opt}) can be iteratively computed by solving the following convex subproblem
\begin{subequations}\label{sub-pro}
\begin{align}
&\mathop {{\rm{maximize}}}\limits_{{\mathbf{W}_k}}~~\frac{1}{{MN_t}}{\sum\limits_{n = 1}^{N_t}}{\sum\limits_{m=1}^M}f(\mathbf{W}_k,\mathbf{W}_{k-1})
\\&{\rm{subject~to}}:~~{\rm{tr}}(\mathbf{W}_k) \le N_t
\\&~~~~~~~~~~~~~~~~~{\mathbf{W}_k} \succeq 0,
\end{align}
\end{subequations}
with
\begin{align}\label{f_Q}
&f(\mathbf{W}_k,\mathbf{W}_{k-1})\\&={f_1}\left( {\mathbf{W}_{k-1}} \right) + {\rm{tr}}\left[ {\nabla{f_1}\left( {{\mathbf{W}_{k-1}}} \right)\left( {{\mathbf{W}_k} - {\mathbf{W}_{k-1}}} \right)} \right]-{f_2}\left( {{\mathbf{W}_k}} \right)\nonumber
\end{align}
where $f(\mathbf{W}_k,\mathbf{W}_{k-1})$ is in form of difference of linear function and convex function, thus it is a concave function with respect to $\mathbf{W}_k$. Therefore, the problem (\ref{sub-pro}) is a convex semi-definite programming (SDP) problem, which can be handled conveniently and efficiently by using the convex optimization toolbox CVX or interior-point method. Finally, the detailed procedures of the proposed Max-ASR-SCA method are presented in Algorithm \ref{Alg2}.
\begin{algorithm}[htp]
\caption{The proposed Max-ASR-SCA method for solving problem (\ref{sdr-opt})}\label{Alg2}
\begin{algorithmic}[1]
\STATE Initialization: find a feasible point $\mathbf{v}_0$, $\mathbf{W}_0=\mathbf{v}_0{\mathbf{v}_0}^H$, with satisfying the constraints in problem (\ref{sdr-opt}), set the tolerance $\epsilon$, and $k=0$
\STATE $\rm{\mathbf{Repeat}}$
\STATE ~~~~Solve the problem $(\ref{sub-pro})$ with $\mathbf{W}_k$, and obtain $\mathbf{W}_k^*$
\STATE ~~~~Set $k\leftarrow k+1$
\STATE ~~~~Update $\mathbf{W}_k=\mathbf{W}_k^*$
\STATE ~~~~Calculate ${R_s'}(\mathbf{W}_k)$ by using (\ref{re-obj})
\STATE $\rm{\mathbf{Until}}$ $\mid{R_{s}'(\mathbf{W}_k)}-{R_{s}'(\mathbf{W}_{k-1})}\mid \leq \epsilon$ is met
\STATE Output the optimal $\mathbf{W}^*=\mathbf{W}_{k-1}$
\end{algorithmic}
\end{algorithm}

It is worth noting that as long as the each approximation of objective function satisfies the tightness and differentiation conditions, the proposed Max-ASR-SCA method is guaranteed to converge to a point that satisfies the Karush-Kuhn-Tucker (KKT) optimality conditions of problem (\ref{sdr-opt}) \cite{Mehanna2014Feasible}. Meanwhile, we present a simple proof for the convergence of the Max-ASR-SCA method in Appendix A.

By using the proposed Max-ASR-SCA method, we actually obtain a SDR solution $\mathbf{W}^*$ to problem (\ref{sdp-opt}). It is worthwhile noting that the solution is an approximately optimal solution. If the solution $\mathbf{W}^*$ is rank-one, the optimal $\mathbf{v}^*$ is the eigenvector corresponding to the largest eigenvalue of $\mathbf{W}^*$. In fact, the solution may not be rank one, we may employ the randomization method to search an approximated solution. Such a randomization method has be empirically found to provide efficient approximations for a variety of applications \cite{Luo2010Semidefinite}. Until now, we complete our design of the proposed two methods.

\subsection{Complexity Analysis and Comparison}
In this subsection, we conduct the complexity analysis on our  proposed two linear precoding methods and compare them with the complexity of the GD method in \cite{Aghdam2016Physical}. Here, consider that a matrix-vector multiplication $y=\mathbf{A}\mathbf{x}$, where $\mathbf{A}\in \mathbb{C}^{m\times n}$ costs $2mn$ floating-point operations (FLOPs), and a matrix-matrix product $\mathbf{C}=\mathbf{A}\mathbf{B}$, where $\mathbf{A}\in \mathbb{C}^{m\times n}$ and $\mathbf{B}\in \mathbb{C}^{n\times p}$, costs $2mnp$ FLOPs (See Appendix C in \cite{Convex} for details).

For the proposed GD method in subsection A, its computational complexity depends  mainly on the following three parts: (a) compute the approximation of SR, (b) compute the gradient of ASR, and (c) the number of iterations. After the complexity of exponential and logarithm operations in (\ref{mi-bob-lb}) and (\ref{mi-eve-lb}) are omitted, we have the computational complexity of part (a)
\begin{align}\label{comp_a}
\mathcal{C}_a=2M^2N_t^2(4N_t^2+N_b+N_e).
\end{align}
In (\ref{grad}), computing the terms $\mathbf{E}_{n,m}^{n',m'}$ and $\mathbf{B}_{n,m}^{n',m'}$ both require about $3N_t^2$ FLOPs, where we ignore the complexity of computing the ${\mathbf{H}}^H\mathbf{Q}_b^{-H}{\mathbf{H}}$ and ${\mathbf{G}}^H\mathbf{Q}_e^{-H}{\mathbf{G}}$, which are invariant matrices for a specifical channel. In fact, $\mathbf{\Delta}_{n,m}^{n',m'}$ is a sparse matrix, which has at most four non-zero elements. Therefore, the complexity of part (b) may be approximately given by
\begin{align}\label{comp_b}
\mathcal{C}_b=2M^2N_t^2(3N_t^2+2N_t^2+2N_t+2N_t^2).
\end{align}
Let $D_1$ denotes the number of iterations in Algorithm \ref{Alg1}, then the final computational complexity of the Max-ASR-GD method is  as follows
\begin{align}\label{comp_1}
\mathcal{C}_{{\rm{Alg}}.1}=2D_1M^2N_t^2(11N_t^2+2N_t+N_b+N_e).
\end{align}

For the Max-SR-GD method in \cite{Aghdam2016Physical}, its complexity analysis is similar to (\ref{comp_1}). Let $N_{\rm{samp}}$ denote the number of realizations of noise for accurately evaluating the actual SR,  then the computational complexity of evaluating SR is about
\begin{align}\label{comp_2a}
\mathcal{C}_e=4{M^2}{N_t}^2{N_{samp}}\left( {2{N_t}^2 + {N_b} + {N_e}} \right).
\end{align}
It is worthwhile noting that the actual SR in (\ref{mi-bob}) requires averaging over at least $N_{\rm{samp}}=500$ realizations of noise for accurately evaluating SR. Similarly, the computational complexity of evaluating the current gradient vector is
\begin{align}\label{comp_2b}
\mathcal{C}_g=2M^2N_t^2N_{samp}(3N_t^2+2N_t^2+2N_t+2N_t^2).
\end{align}
Let $D_2$ denote the number of iterations for Algorithm.1 in \cite{Aghdam2016Physical}, then its complexity is written as
\begin{align}\label{comp_2}
\mathcal{C}=2D_2M^2N_t^2N_{samp}(11N_t^2+2N_t+2N_b+2N_e).
\end{align}

Finally, for the proposed Max-ASR-SCA method, its computational complexity also depends on the following three aspects: (a) compute the (\ref{re-obj}), (b) solve the convex optimization problem in (\ref{sub-pro}), and (c) the number of iterations. The computational complexity of aspect (a) is about $2{M^2}{N_t}^2\left( {2N_t^3 + 3{N_t}^2} \right)$ FLOPs. It is noted that the exact complexity of aspect (b) depends heavily on the specific convex solver. Here, it is assumed that the interior point method \cite{Helmberg1996An} is used. The objective function of problem (\ref{sub-pro}) consists of $MN_t$ convex functions, one linear inequality constraint, and one linear matrix inequality constraint. According to \cite{Luo2010Semidefinite,Helmberg1996An}, this problem in (\ref{sub-pro}) may be solved with a worst-case complexity
\begin{align}\label{comp_sdp}
\mathcal{C}_{{\rm{sdp}}}=\mathcal{O}(N_t^{4.5}\log(1/\varsigma))+2M^2N_t^2(3N_t^3+4N_t^2),
\end{align}
where $\varsigma$ is a given solution accuracy, and the second term is the complexity of constructing the objective function. Let $D_3$ denotes the number of iterations in Algorithm \ref{Alg2}. Therefore, the complexity of the proposed Max-ASR-SCA method in Algorithm.\ref{Alg2} is approximated as
\begin{align}\label{comp_sca}
\mathcal{C}_{{\rm{Alg}}.2}=D_3\mathcal{C}_{{\rm{sdp}}}+2 D_3{M^2}{N_t}^2\left( {2N_t^3 + 3N_t^2} \right).
\end{align}

In summary, from the above complexity analysis, it can be seen that the three methods all have a polynomial complexity. The proposed Max-ASR-GD in Algorithm.~\ref{Alg1} has the lowest complexity among the three schemes, and the proposed Max-ASR-SCA method is in between Max-ASR-GD and Max-SR-GD in \cite{Aghdam2016Physical} in terms of complexity due to $N_{samp}\gg N_t$. This means that Max-SR-GD is of the highest complexity among the three methods. From (\ref{comp_1}) and (\ref{comp_sca}), it follows that the complexities of Max-ASR-GD and Max-ASR-SCA have the orders  $\mathcal{O}(N_t^4)$ and $\mathcal{O}(N_t^5)$ provided that other parameters are fixed, respectively. Therefore, when the number of transmit antennas tends to large-scale, the proposed Max-ASR-GD method has resulted in a dramatic reduction on computational complexity compared to Max-ASR-SCA and Max-SR-GD.

\section{Numerical Results and Discussions}
In this section, we present our numerical simulation results to evaluate the proposed two linear precoding methods. In the following simulations, the noise power of Bob and Eve is assumed to be equal, i.e., $\sigma_b^2=\sigma_e^2$, and the achievable SR is evaluated by using (\ref{ave-sr}) and calculated by averaging over $50$ realizations of channels. Meanwhile, some necessary parameters are set as follows: $N_b=N_e=2$; $\mu=0.5$, and $\mu_{\rm{min}}=0.01$ for Algorithm \ref{Alg1}; $\epsilon=0.001$ for Algorithm \ref{Alg2}.

\begin{figure}
 \centering
 \includegraphics[width=3.4in, height=2.7in]{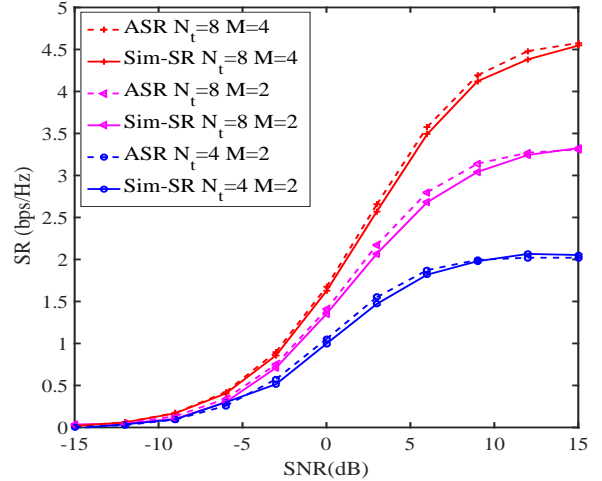}\\
 \caption{Approximated and simulated SR}\label{sr_gap}
\end{figure}

First, in order to verify the validiness of using ASR as the design metric. Fig.~\ref{sr_gap} plots the curves of the approximated and simulated actual SRs versus signal-to-noise ratio (SNR), where Sim-SR stands for the simulated actual SR for the sake of simplicity. It is noted that the approximated and actual SR are calculated by (\ref{app-sr}) and (\ref{Rs}), respectively. It can be seen that all ASR curves are very close to these of simulated actual SR for almost all SNR regions. In other words, the rate difference between ASR and simulated actual SR is trivial. Therefore, it is reasonable to employ the closed-form expression of ASR in (\ref{app-sr}) as the design metric.

\begin{figure}
 \centering
 \includegraphics[width=3.4in, height=2.7in]{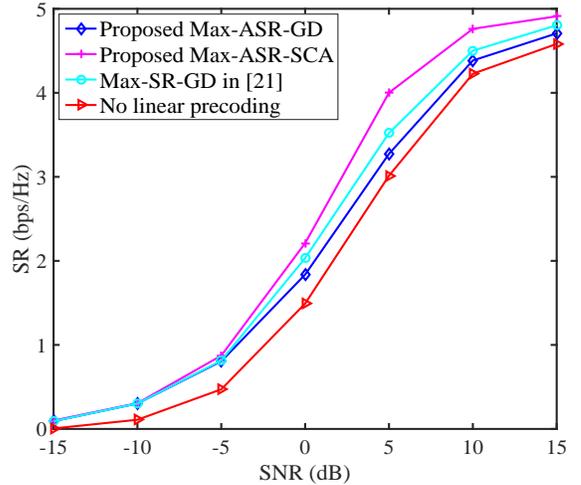}\\
 \caption{SR performance with $N_t=8$ and $M=4.$}\label{sr_qpsk}
\end{figure}

Fig.~\ref{sr_qpsk} demonstrates  the  curves of SR performance versus SNR for three linear precoding methods Max-ASR-GD, Max-ASR-SCA, and Max-SR-GD with no linear precoding scheme as a performance benchmark. It can be seen that the three methods can achieve  different  SR performance gains over the no precoding case. The SR performance of proposed Max-ASR-GD method is worse than Max-ASR-SCA but tends to conventional Max-SR-GD  in \cite{Aghdam2016Physical}. It has a high probability that the two based-GD methods, Max-ASR-GD and Max-SR-GD, converge to their locally optimal solutions while the proposed Max-ASR-SCA method can obtain the approximately optimal solution. Therefore, the proposed Max-ASR-SCA achieves the best SR performance among the three precoding schemes.

\begin{figure}
 \centering
 \includegraphics[width=3.4in, height=2.7in]{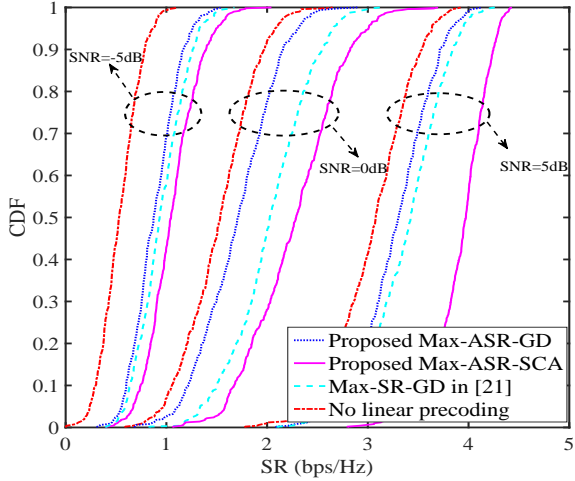}\\
 \caption{CDF of SR with $N_t=8$ and $M=4.$}\label{cdf_qpsk}
\end{figure}

Fig.~\ref{cdf_qpsk} shows the cumulative distribution function (CDF) curves of SR for the three linear precoding methods with different typical values of SNRs: -5dB, 0dB, and 5dB. The curves are generated by calculating SR over 500 realizations of all channels. For all three methods, as SNR increases, the CDF curves of SR moves towards the right-hand side. This means that the value of SR becomes large as SNR increases. Additionally, for all three given values of SNR, the four methods have a decreasing order in terms of SR performance as follows: Max-ASR-SCA, Max-SR-GD in \cite{Aghdam2016Physical}, Max-ASR-GD, and without linear precoding.

\begin{figure}
 \centering
 \includegraphics[width=3.4in, height=2.7in]{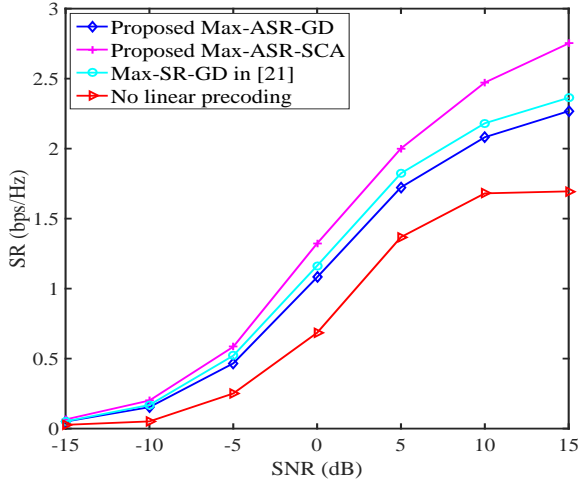}\\
 \caption{SR performance with $N_t=4$ and $M=2.$}\label{sr_bpsk}
\end{figure}

In order to examine the impact of $N_t$ and $M$ on SR performance, Fig.~\ref{sr_bpsk}  presents the curves of SR performance versus SNR for the three methods with $N_t=4$,~ and $M=2$. It can be seen that the the SR performance trend in Fig.~\ref{sr_bpsk} is similar to that in Fig.~\ref{sr_qpsk}. Compared to no linear precoding, the proposed Max-ASR-SCA and Max-ASR-GD, and the Max-SR-GD method in \cite{Aghdam2016Physical} achieve 1 bps/Hz, 0.7 bps/Hz, and 0.6 bps/Hz SR improvement at SNR=15dB, respectively. Compared to conventional Max-SR-GD, the SR improvement gain achievable by the proposed Max-ASR-SCA is about 13\% at SNR=15dB, which is a nice result.

\begin{figure}
 \centering
 \includegraphics[width=3.4in, height=4.5in]{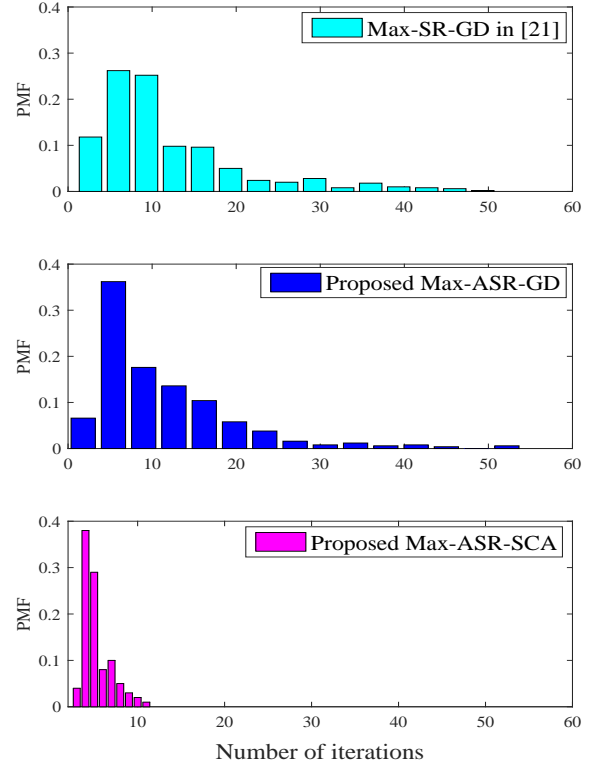}\\
 \caption{PMF of numbers of iterations  with $N_t=4,~M=2$, and SNR$=5$dB.}\label{pmf}
\end{figure}

Fig.~\ref{pmf} illustrates the histogram of probability mass function (PMF) of number of iterations for the proposed Max-ASR-SCA, and Max-ASR GD with conventional Max-SR-GD in \cite{Aghdam2016Physical} as a performance reference. It is noted that 500 channel trials are conducted to show the distribution of number of iterations. It can be seen that near 90 percent trials converge within 30, 25 and 8 iterations for conventional Max-SR-GD in \cite{Aghdam2016Physical}, the proposed Max-ASR-GD, and Max-ASR-SCA methods, respectively. In other words, the proposed Max-ASR-GD converges slightly faster than conventional Max-SR-GD but slower than proposed Max-ASR-SCA. Although the proposed Max-ASR-SCA method converges within a fewer number of iterations, its complexity of each iteration is higher than that of the proposed Max-ASR-GD method.

\begin{figure}
 \centering
 \includegraphics[width=3.4in, height=2.7in]{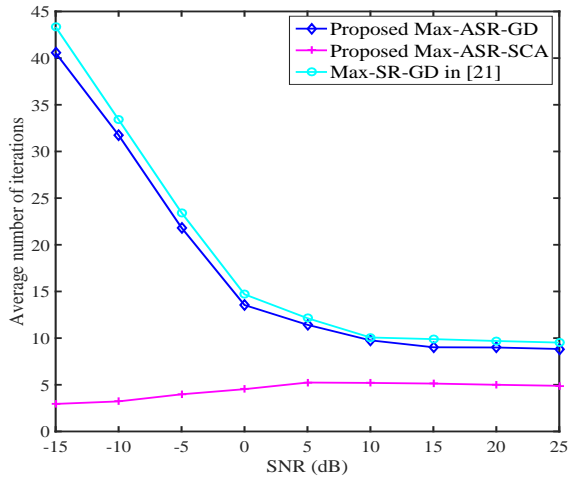}\\
 \caption{Average number of iterations versu SNR with $N_t=4$ and $M=2$.}\label{iter_snr}
\end{figure}

Fig.~\ref{iter_snr} presents the curves of the average number of iterations versus SNR for the above three linear precoding schemes. It is observed that the average number of iterations of for the two based-GD methods decrease as SNR increases. However, for the Max-ASR-SCA scheme, the average number of iterations slightly increases when SNR increases. When SNR tends to be larger than 15dB, the average number of iterations nearly keep constant for three linear precoding schemes. From Fig.~\ref{iter_snr}, it also can be seen that the three schemes have an increasing order in terms of convergence performances as follows: Max-SR-GD, Max-ASR-GD, and Max-ASR-SCA.

\begin{figure}
 \centering
 \includegraphics[width=3.4in, height=2.7in]{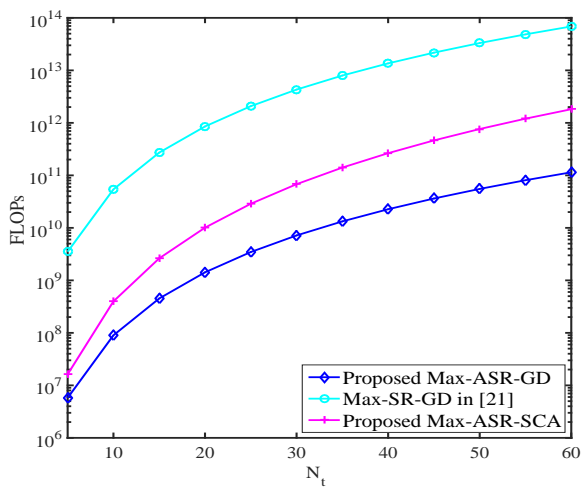}\\
 \caption{Numbers of FLOPs versu $N_t$ with $M=4.$}\label{complexity}
\end{figure}

To make a detailed complexity comparison among the three linear precoding schemes, Fig.~\ref{complexity} shows the curves of their complexities versus $N_t$. In accordance with Fig.~\ref{pmf}, the numbers of iterations are set as $D_1=25,D_2=30,D_3=8$ for insuring 90\% trials converge. It can be seen that the Max-SR-GD method in \cite{Aghdam2016Physical} has an extremely high complexity compared to the two proposed methods. The complexities of the proposed Max-ASR-GD and Max-ASR-SCA are two-order-of-magnitude, and one-order-of-magnitude lower than that of conventional Max-SR-GD as the value of $N_t$ is larger than 30, respectively. Thus, the proposed two methods leads to a significant complexity reduction over Max-SR-GD. This makes them applicable to the future secure spatial modulation networks. Furthermore, compared to Max-SR-GD and Max-ASR-SCA, the main advantage of the proposed  Max-ASR-GD is its extremely-low-complexity. Especially, as the value of $N_t$ tends to medium-scale and large-scale, its computational complexity is far lower than those of Max-SR-GD and Max-ASR-SCA.

\section{Conclusion}
In this paper, we designed two linear precoding schemes for secure SM systems. In order to simplify the  original optimization problem of Max-SR-GD and reduce its computational complexity, a simple and analytical expression of ASR for SR was derived as a design or optimization metric. Subsequently, two low-complexity linear precoding schemes, namely Max-ASR-GD and Max-ASR-SCA, were developed based on convex optimization techniques. Our examination first demonstrated that the proposed Max-ASR-SCA outperforms the conventional Max-SR-GD in terms of achieving a higher SR with a one-order-of-magnitude lower complexity. In addition, the proposed Max-ASR-GD achieves a two-order-of-magnitude lower complexity than Max-SR-GD at the cost of a negligible SR performance loss, which shows that the proposed Max-ASR-GD strikes a good balance between the complexity and SR performance. The proposed two low-complexity linear precoders are attractive for future MIMO networks with secure spatial modulation.

\begin{appendices}
\section{Proof of the convergence for the Max-ASR-SCA}
First, consider that $\mathbf{W}_{k}$ is a feasible solution of problem (\ref{sdr-opt}) and $f_1(\mathbf{W})$ is a convex function with respect to $\mathbf{W}$, so we have the following inequality
\begin{align}\label{taylor-k}
{f_1}\left( {{\mathbf{W}_{k+1}}} \right) \ge {f_1}\left( {{\mathbf{W}_k}} \right) + {\rm{tr}}{\left( \nabla{f_1}\left( {{\mathbf{W}_k}} \right)\left( {{\mathbf{W}_{k+1}} - {\mathbf{W}_k}} \right) \right)},
\end{align}
where $\mathbf{W}_{k+1}$ is a updated solution of problem (\ref{sdr-opt}) by solving the problem (\ref{sub-pro}) with a given feasible point $\mathbf{W}_{k}$. Then, according to the objective function in (\ref{sub-pro}a), we have the following inequality
\begin{align}\label{prof-ineq}
{\sum\limits_{n = 1}^{N_t}}{\sum\limits_{m=1}^M}f(\mathbf{W}_{k+1},\mathbf{W}_k) \geq {\sum\limits_{n = 1}^{N_t}}{\sum\limits_{m=1}^M}f(\mathbf{W}_k,\mathbf{W}_k).
\end{align}
Based on the definition of $f(\mathbf{W},\mathbf{W}_0)$ in (\ref{f_Q}), $f(\mathbf{W}_{k+1},\mathbf{W}_k)$ and $f(\mathbf{W}_k,\mathbf{W}_k)$ is given by
\begin{align}\label{f_W}
&f(\mathbf{W}_{k+1},\mathbf{W}_k)\\&={f_1}\left( {\mathbf{W}_k} \right) + {\rm{tr}}\left[ {\nabla{f_1}\left( {{\mathbf{W}_k}} \right)\left( {\mathbf{W}_{k+1} - {\mathbf{W}_k}} \right)} \right]-{f_2}\left( {\mathbf{W}_{k+1}} \right)\nonumber,
\end{align}
\begin{align}\label{f_W2}
f(\mathbf{W}_k,\mathbf{W}_k)={f_1}\left( {\mathbf{W}_k} \right)-{f_2}\left( {{\mathbf{W}_k}} \right).
\end{align}
Finally, by substituting (\ref{f_W}) and (\ref{f_W2}) into (\ref{prof-ineq}) and using the inequality in (\ref{taylor-k}), we have the following inequality
\begin{align}\label{last-ineq}
&{\sum\limits_{n = 1}^{N_t}}{\sum\limits_{m=1}^M}\left[{f_1}\left( {{\mathbf{W}_{k+1}}} \right)-{f_2}\left( {{\mathbf{W}_{k+1}}} \right)\right] \nonumber\\&\geq {\sum\limits_{n = 1}^{N_t}}{\sum\limits_{m=1}^M} \left[{f_1}\left( {{\mathbf{W}_k}} \right)-{f_2}\left( {{\mathbf{W}_k}} \right)\right].
\end{align}

It is observed the achieved objective function value in (\ref{sdr-opt}) is monotone non-decreasing and bounded as the number of iterations increases, thus we may conclude the SCA algorithm converges. Until now, we complete the proof of the convergence for the Max-ASR-SCA.
\end{appendices}

\bibliographystyle{IEEEtran}
\bibliography{IEEEabrv,lp_ssm}

% that's all folks
\end{document}